\documentclass[12pt]{article}

\RequirePackage{titlesec}
\usepackage{graphicx}
\usepackage{amssymb}
\usepackage{amsmath}
\usepackage{hyperref}

\titleformat{\section}{\bf\Large\raggedright}{}{0em}{}

\begin{document}

\thispagestyle{empty}

\begin{center}
${}$\\
{ \Large \bf Causal Dynamical Triangulations:
\\ \vspace{10pt} New Lattice Theory of Quantum Gravity\footnote{Invited contribution to Scholarpedia}
}

\vspace{18pt}

{\sl J. Ambj\o rn}$\,^{a,b}$
and
{\sl R. Loll}$\,^{a}$

\vspace{4pt}

{\footnotesize
$^a$~Institute for Mathematics, Astrophysics and Particle Physics, Radboud University \\ 
Heyendaalseweg 135, 6525 AJ Nijmegen, The Netherlands.\\ 
{email: r.loll@science.ru.nl}\\

\vspace{6pt}

$^b$~The Niels Bohr Institute, Copenhagen University\\
Blegdamsvej 17, DK-2100 Copenhagen \O , Denmark.\\
{email: ambjorn@nbi.dk}\\
}
\end{center}

\vspace{4pt}

\begin{center}
\textbf{Abstract}
\end{center}
\noindent
Causal Dynamical Triangulations (CDT) is a methodology to define and compute the gravitational path integral, whose aim is a fully fledged non\-per\-tur\-bative quantum field theory of gravity and spacetime. Analogous to lattice formulations of nongravitational quantum fields, CDT provides a blueprint for lattice quantum gravity, where -- crucially -- the dynamical, curved and causal nature of spacetime is built into the structure of the lattices from the outset. The regularized path integral involves a sum over triangulated spacetimes, each assembled from flat, Minkowskian building blocks. The degrees of freedom of general relativity are encoded in a coordinate-free manner in the neighbourhood relations of the building blocks and the length of their edges, which also serves as a short-distance cutoff. 

A well-defined Wick rotation makes this path integral amenable to Monte Carlo simulations. Despite the absence of an a priori preferred background geometry, numerical experiments have revealed the dynamical emergence of a quantum universe near the Planck scale. Its global properties are compatible with those of a de Sitter space, providing strong evidence for a well-defined classical limit. At the same time, large quantum fluctuations lead to unex\-pec\-ted properties on short scales, most prominently, a spectral dimen\-sion near 2, replacing the classical value of 4. Computer simulations indicate the presence of an ultraviolet fixed point under renormalization, opening the door to a nontrivial continuum theory. Efforts are under way to construct observables that can elucidate the nonperturbative quantum origins of early-universe cosmology.    
${}$\\
\vspace{2cm}

\newpage

\tableofcontents

\newpage

\addcontentsline{toc}{section}{1\hspace{0.3cm} Laying the ground}
\section*{1\hspace{0.2cm} Laying the ground}

Formulating quantum gravity in terms of \textbf{Causal Dynamical Triangulations (CDT)} is grounded in the well-known ingredients and principles of general relativity and
quantum field theory, combining them in new, intrinsic ways \cite{Ambjorn1998,Ambjorn2001}. 
In a nutshell, \textit{classical gravity} is the field theory of Lorentzian metrics $g_{\mu\nu}$ on a four-dimensional spacetime $M$ with the Einstein-Hilbert action 
\begin{equation}
\label{eq:contact}
S[g_{\mu\nu};G,\Lambda] =
 \frac{ 1}{16 \pi G}\int_M d^4 x \sqrt{-g(x)} \, \big( R(x) - 2 \Lambda\big),
 \end{equation}
where $g$ is the determinant of $g_{\mu\nu}$, $R$ its scalar curvature, 
$G$ the gravitational and $\Lambda$ the cosmological constant.
This article uses units where $c \!=\! \hbar \! =\! 1$. 
 
The corresponding \textit{quantum field theory of gravity} is defined formally by the path inte\-gral
\begin{equation}
\label{eq:pi}
Z(G,\Lambda) = \int {\cal D} [g_{\mu\nu}] \; e^{iS[g_{\mu\mu};G,\Lambda]} 
\end{equation}
where the integration is over four-geometries $ [g_{\mu\nu}]$, i.e.\ metrics modulo four-diffeomorphisms. 
It is not clear a priori which geometries should be included in the path integral, but in analogy with nongravitational quantum field theo\-ries 
one would expect it to contain at least all continuous four-geometries of a manifold $M$ of fixed topology. It is 
not known how to include a sum over different topologies, since these cannot even be classified.

The formal character of the path integral (\ref{eq:pi}) cannot be addressed by expanding $Z$ as a perturbative series in $G$,
based on the field decomposition 
\begin{equation}
g_{\mu\nu} = g_{\mu\nu}^\mathrm{bg} + \sqrt{G} \, h_{\mu\nu}
\end{equation}
into a fixed background metric $g_{\mu\nu}^\mathrm{bg}$ (for $\Lambda\! =\! 0$, the Minkowski metric $\eta_{\mu\nu}$) and fluctuations $h_{\mu\nu}(x)$,
because this perturbation theory is not renormalizable. 
It calls for a \textit{nonperturbative definition of} $Z$, which is exactly what CDT provides. 
CDT is a lattice implementation of $Z(G,\Lambda)$, where the dynamical and nonperturbative nature of geometry is built in from the outset. 
It has a short-distance,  ultraviolet (UV) cutoff, given by the length $a$ of a lattice edge, which
makes the regularized lattice theory well defined. To obtain its con\-tinuum limit 
one investigates the limit $a\! \to\! 0$, as will be described below. 

\addcontentsline{toc}{section}{2\hspace{0.3cm} Constructing curved, causal lattices}
\section*{2\hspace{0.2cm} Constructing curved, causal lattices}

Building on earlier attempts to put quantum gravity on the lattice \cite{Loll1998}, which have not found interesting results,
the distinguishing features of CDT's lattices are their \textit{intrinsic curvature} and \textit{causal structure}. 
The configuration space is \textit{not} given by a fixed, rigid (hypercubic or other) lattice with variable assignments of metric 
quantities to its lattice elements, but consists of variable ``gluings'' of a fixed number of just two types of identical geometric building blocks.
While each simplicial (=triangular) building block is geometrically a piece of flat Minkowski space, how they are assembled 
into four-dimensional triangulated manifolds encodes nonvanishing intrinsic curvature and a lattice analogue of global hyperbolicity, i.e.\ a well-defined causal
structure.  

It is important to understand that most details of how building blocks are chosen and assembled into path integral configurations will -- by the mechanism
of \textit{universality} -- make no difference to the final continuum theory, if it exists. Research up to now has found evidence for the existence of two
distinct universality classes for nonperturbative constructions of quantum gravity: the causal, Lorentzian one of CDT discussed here, and another one, 
associated with the purely Euclidean
precursor of CDT \cite{Ambjorn2024}, whose partition function seems to be dominated by pathological, unphysical configurations. 

To implement global hyperbolicity, each CDT configuration
has the form of a sequence of three-dimensional spatial triangulations $\Sigma (t)$, each labelled by an integer $t$ that measures the number of discrete proper-time
steps from some initial configuration $\Sigma(0)$. Like in canonical quantum gravity, the spatial slices are not allowed to change topology as a function of time,
implying a product topology $[0,1]\times\Sigma$ for all path integral histories. In what follows, the spatial topology will be that of a three-sphere $S^3$, unless stated otherwise.

\begin{figure}[t]
\centerline{\scalebox{0.5}{\rotatebox{0}{\includegraphics{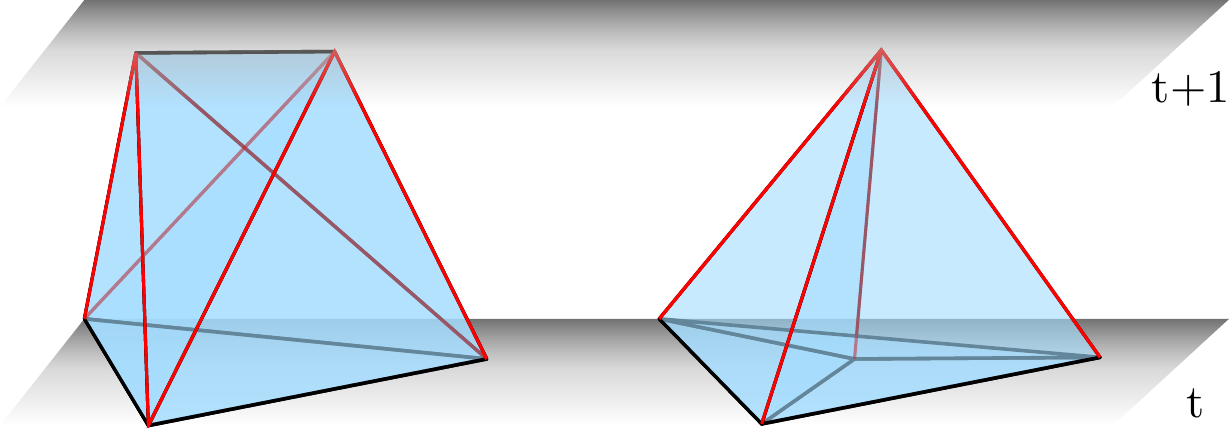}}}}
\vspace{0.4cm}
\caption{ \small Elementary Minkowskian building blocks of CDT, of type (3,2) (left) and type (4,1) (right), and their position
inside a spacetime slice $[t,t\! +\! 1]$. Timelike edges (red) interpolate between $t$ and $t\!+\! 1$, spacelike ones (black) have a fixed $t$.
}
\label{fig-am2}
\end{figure}

The two types of constituent four-simplices of CDT lattices are shown in Fig.\ \ref{fig-am2}. 
Since by construction all lattice vertices lie in spatial slices of integer time, a four-simplex type is characterized by a pair $(m,n)$ whenever $m$ ($n$) of its vertices have
time label $t$ ($t\!+\! 1$). Together with the (3,2)- and (4,1)-simplices shown, also the time-reflected (2,3)- and (1,4)-simplices occur. 
All of them have spacelike edges of squared length $a^2$ and
timelike edges of squared length $-\alpha a^2$, for some positive constant $\alpha$. 
Each spacelike edge is contained in a spatial triangulation $\Sigma(t)$, while timelike edges interpolate between
adjacent three-geometries $\Sigma(t)$ and $\Sigma (t\! +\!1)$. Neighbouring four-simplices share three-dimensional tetrahedral faces and fill out all spacetime slices 
$[t,t+1]$.  

The configuration space of the lattice-regularized gravitational path integral
consists of all geometrically distinct, piecewise flat Lorentzian spacetimes obtained by assembling 
the elementary building blocks according to the causal gluing rules sketched above. 
Note that these geometries are continuous, but not smooth, with singular curvature assignments along
their two-dimensional subsimplices \cite{review1}. 
Assuming a finite number of time steps and building blocks, the
number of different simplicial manifolds that can be obtained in this way, subject to chosen boundary conditions, is \textit{finite}.  

\addcontentsline{toc}{section}{3\hspace{0.3cm} Gravitational action and analytic continuation}
\section*{3\hspace{0.2cm} Gravitational action and analytic continuation}
  
Apart from a configuration space, a lattice version of the path integral (\ref{eq:pi}) requires
a lattice implementation of the continuum action (\ref{eq:contact}). Since CDT con\-fi\-gu\-rations are
a special class of piecewise flat simplicial manifolds, one can follow Regge's prescription \cite{Regge} to
express the Einstein-Hilbert action as a function of the edge lengths and lattice connectivity.\footnote{This
has the character of a finite-difference expression and is neither `exact' nor unique.} 
Also needed is an analytic continuation to allow for an explicit computation of the path integral and
a controlled evaluation of its continuum limit. A key ingredient of CDT is a well-defined
\textit{Wick rotation}, defined by continuing the parameter $\alpha$ to $-\alpha$ in the lower-half complex $\alpha$-plane, 
which renders the path integral real \cite{review1,review2}. It maps timelike edge lengths $-\alpha a^2$ to
spacelike ones $|\alpha | a^2$, and therefore associates with each Lorentzian CDT \textit{spacetime} a unique
Riemannian tri\-an\-gu\-lated \textit{space}, without affecting the underlying abstract triangulation $T$.  

Choosing $\alpha\! =\! 1$ for definiteness, the analytic continuation of the Lorentzian Regge action $S_L$
gives ($i$ times) the corresponding Euclidean action $S_E$, 
 \begin{equation}
 S_L[T,\alpha \!=\! 1] \to  \lim_{\epsilon \to 0} S_L[T, \alpha\! =\! -1- \! i\epsilon] = i S_E[T].
 \end{equation}
Since after Wick-rotating all triangulations are equilateral (with edge length $a$), the functional form of the action
becomes exceedingly simple,  
\begin{equation}
\label{eq:euact}
S_E[T] = -\hat{k}_0 N_0(T) + \hat{k}_4 N_4(T),
\end{equation}
where $N_0(T)$ and $N_4(T)$ denote the numbers of vertices and four-simplices
in $T$.\footnote{A term proportional to the Euler characteristic of $T$ has been dropped, because it is
negligible for large $N_i(T)$.}
The dimensionless real lattice coupling constants $\hat{k}_0$ and $\hat{k}_4$ are related to the 
couplings $G$ and $\Lambda$ of the continuum action (\ref{eq:contact}) by
\begin{equation}
\hat{k}_0 =  c_0\, a^2/G, \qquad \hat{k}_4 = c_4\, a^2/G + c_4' \, a^4 \Lambda/G, 
\end{equation}
where $c_0$, $c_4$ and $c'_4$ are positive constants of order 1.  
The existence of a well-defined Wick rotation beyond perturbation theory is unique in the context of quantum gravity,
and makes the CDT formulation amenable to Monte Carlo simulations.

\addcontentsline{toc}{section}{4\hspace{0.3cm} CDT implementation of the path integral}
\section*{4\hspace{0.2cm} CDT implementation of the path integral}

With these ingredients, the CDT lattice implementation of the path integral (\ref{eq:pi}) is
\begin{equation}
\label{eq:cdtpi}
Z^{L} = \sum_{T} \text{\footnotesize  $\frac{1}{C_T}$} \, e^{i S_L[T;\alpha =1]}, 
\end{equation}
where the sum is taken over distinct, unlabelled causal triangulations as introduced above and $C_T$ is a symmetry factor 
(the order of the automorphism group of $T$, which for large $T$ is almost always equal to 1).
The absence of labels, e.g.\ for the vertices or four-simplices of the triangulations, reflects the fact that the
path integral does not suffer from any coordinate (=rela\-bel\-ling) redundancies.
The right-hand side of (\ref{eq:cdtpi}) only becomes computationally tractable after analytic continuation,
which yields the real partition function (Euclidean path integral)
\begin{equation}
\label{eq:eupi}
Z^E(\hat{k}_0,\hat{k}_4) = \sum_{T} \text{\footnotesize $\frac{1}{C_T}$} \, e^{-S_E[T;\hat{k}_0,\hat{k}_4]},
\end{equation}
depending now on the Euclidean action $S_E$ of eq.\ (\ref{eq:euact}). 
Each triangulation $T$ contributes with a positive Boltzmann factor $\exp (-S_E[T])$,   
and the associated statistical system can be studied by using Monte Carlo simulations \cite{review1}.

Note that the construction just sketched entails a loss of generality: while 
the Lorentzian Regge action depended in a different way on the (3,2)-simplices and
the (4,1)-simplices, due to the different geometry of the two types of building blocks,
this is no longer the case for its Euclidean counterpart (\ref{eq:euact}).
This can be traced back to the special choice $\alpha\! =\! 1$, and can be rectified by either
restoring $\alpha$ as a free parameter or, equivalently, allowing for two different coupling constants
$k_{32}$ and $k_{41}$, such that the Euclidean action reads
\begin{equation}
\label{eq:seu1}
S_E[T;k_0,k_{32},k_{41}] = -k_0 N_0(T) + k_{32} N_4^{(3,2)}(T) + k_{41} N_4^{(4,1)}(T),
\end{equation}
where $N_4^{(3,2)}(T)$ counts (3,2)- and (2,3)-simplices 
and $N_4^{(4,1)}(T)$ (4,1)- and (1,4)-simplices in $T$, such that
$N_4^{(3,2)}(T)+N_4^{(4,1)}(T) \! =\! N_4(T)$. A form of the action that
is convenient in Monte Carlo simulations is
\begin{equation}
\label{eq:seu2}
S_E[T;k_0,\Delta,k_4] =  -(k_0 + 6\Delta) N_0(T)  +k_4 N_4(T) + \Delta N_4^{(4,1)},
\end{equation}
which is a rewriting of (\ref{eq:seu1}), making use of so-called Dehn-Sommerville
relations among the numbers of various types of (sub-)simplices of $T$ \cite{review1}.
When the newly introduced $\Delta$-parameter vanishes, $\Delta\! =\! 0$, this reduces to eq.\ (\ref{eq:euact}), with
$k_0\!\rightarrow\! \hat{k}_0$ and $k_4\!\rightarrow\! \hat{k}_4$. 

Finally, the expression for the analytically continued CDT path integral is 
\begin{equation}
\label{eq:newsum}
Z_E (k_0,\Delta,k_4) = \sum_{T} \text{\footnotesize $\frac{1}{C_T}$} \,
e^{-S_E[T;k_0,\Delta,{k}_4]} = \sum_{N_4} e^{-k_4 N_4} Z_E(k_0,\Delta,N_4),
\end{equation}
where in the last step the sum over triangulations $T$ has been decomposed into sums for fixed, discrete
four-volume $N_4$, captured by the partition function 
\begin{equation}
\label{eq:partn4}
Z_E (k_0,\Delta,N_4)\! := \!\!\!\!\!\sum_{T|_{N_4(T) =N_4} } 
\!\!\!\! \text{\footnotesize $\frac{1}{C_{T}}$}\, e^{(k_0+6\Delta)N_0(T) -
\Delta N^{(4,1)}_4 (T)} = 
e^{k_4^c(k_0,\Delta) N_4} {\cal R}(k_0,\Delta,N_4).
\end{equation}
As a function of $N_4$, it grows exponentially as indicated, with a subleading remainder ${\cal R}(k_0,\Delta,N_4)$.
It follows that 
for given values of $k_0$ and $\Delta$, the sum (\ref{eq:newsum}) is convergent for $k_4 > k_4^c(k_0,\Delta)$,
due to the exponential damping of large volumes $N_4$.
In this region of the coupling constant space, spanned by $(k_0,\Delta,k_4)$, the CDT path integral is therefore well defined.
Lastly, it is worth noting the manifestly background-independent and nonperturbative nature of the path integral, which is a 
``democratic'' sum over all causal, curved triangulations, without distinguishing any particular one or requiring them to be close 
to a classical solution of the Einstein equations.

\addcontentsline{toc}{section}{5\hspace{0.3cm} Phase diagram of CDT lattice quantum gravity}
\section*{5\hspace{0.2cm} Phase diagram of CDT lattice quantum gravity}

The above discussion makes it clear that the value chosen for the bare coupling constant $k_4$ 
determines the average discrete volume $\langle N_4\rangle$ in the ensemble of triangulations contributing to the path integral (\ref{eq:newsum})  and 
is largest in the limit that $k_4$ approaches its critical value, $k_4 \!\to\! k_4^c(k_0,\Delta)$, from above. 

However, rather than fine-tuning $k_4$ to $k_4^c(k_0,\Delta)$, whose value is not known analytically,  
in computer simulations it is far more convenient to fix the four-volume $N_4$ and perform calculations for different values of $N_4$.
This means that one uses the ``canonical'' partition function  $Z_E(k_0,\Delta,N_4)$ rather than the ``grand canonical'' $Z_E(k_0,\Delta,k_4)$.
It does not entail any loss of information, since the two partition functions are related by a 
discrete Laplace transformation, as shown by eq.\ (\ref{eq:newsum}). 
 
\begin{figure}[t]
\vspace{-1.5cm}
\centerline{\scalebox{0.7}{\rotatebox{0}{\includegraphics{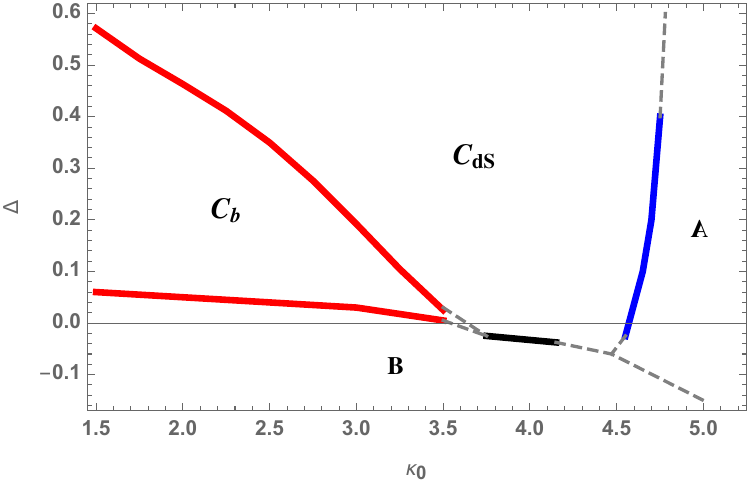}}}}
\vspace{-0.3cm}
\caption{CDT phase diagram, parametrized by the bare coupling constants $k_0$ and $\Delta$ \cite{Ambjorn2020}.
The physically interesting phase is the de Sitter phase $C_{\rm dS}$.}
\label{fig:dia}
\end{figure}

The statistical system associated with $Z_E(k_0,\Delta,N_4)$ has been studied systematically as a function of 
the coupling constants $k_0$ and $\Delta$, revealing several geometrically distinct phases and phase transitions between them
(see Fig.\ \ref{fig:dia}), including phase transitions of second order \cite{Ambjorn2011,Coumbe2015}. Only the so-called de Sitter phase $C_{\rm dS}$
will be of interest here, since it displays a scaling behaviour and observable properties that are compatible with those of a
four-dimensional universe on sufficiently large scales. The other phases appear to be lattice artefacts, which lack any discernible relation
with a continuum theory of gravity. All measurements of observables reported here have been taken in the de Sitter phase. 

The evidence for the emergence of an extended spacetime is a breakthrough result of CDT \cite{Ambjorn2004} and stands out compared to previous attempts at lattice quantum gravity. 
A well-known difficulty of any nonperturbative, back\-ground-independent formulation of quantum gravity is that of the ``classical limit'',
i.e.\ showing that its (sub-)Planckian dynamics and degrees of freedom in a large-distance limit have anything to do with gravity and spacetime as we know it. 
The nature of the evidence for this in lattice CDT is by no means complete but already compelling. It is quantified in terms of suitable \textit{observables},
which characterize the emergent quantum geometry in the de Sitter phase, as will be spelled out in more detail below. 

The construction of diffeomorphism-invariant observables, which are well defined in a Planckian regime, is another major and well-known challenge of quantum gravity.
Suffice it to say that a concrete and functioning computational setting, like that available in CDT lattice gravity, amounts to a step change in the often abstract debate around
observables \cite{Loll2025}: quantum observables can be designed and implemented, their expectation values measured, and the results 
fed back into the further construction of the theory.

\addcontentsline{toc}{section}{6\hspace{0.3cm} Macroscopic quantum de Sitter universe}
\section*{6\hspace{0.2cm} Macroscopic quantum de Sitter universe}

A pivotal observable in the development of the lattice theory has been the \textit{volume profile}, given by 
the expectation value $\langle N_3(i)\rangle$ of the spatial three-volume of spacetime as a function of the integer time $t\! =\! i$, where 
$N_3(i)$ counts the tetrahedra in the triangulated, spatial submanifold at time step $i$. This ``time" does not have an immediate physical meaning, but 
the measurements of the volume profile indicate that in the continuum limit it can be interpreted as (proportional to) global proper time, as will become clear below.
In what follows, time is cyclically identified, such that the global topology of the path integral configurations is $S^1\!\times\! S^3$. 
However, intriguingly, it turns out that the collective dynamics of the microscopic degrees
of freedom drives the system to a quantum universe whose \textit{effective}\footnote{i.e.\ on scales much larger than the lattice cutoff $a$}
topology is that of a four-sphere $S^4$.   

\begin{figure}[t]
\vspace{-1cm}
\hspace{-0.7cm}
\centerline{{\scalebox{0.137}{\includegraphics{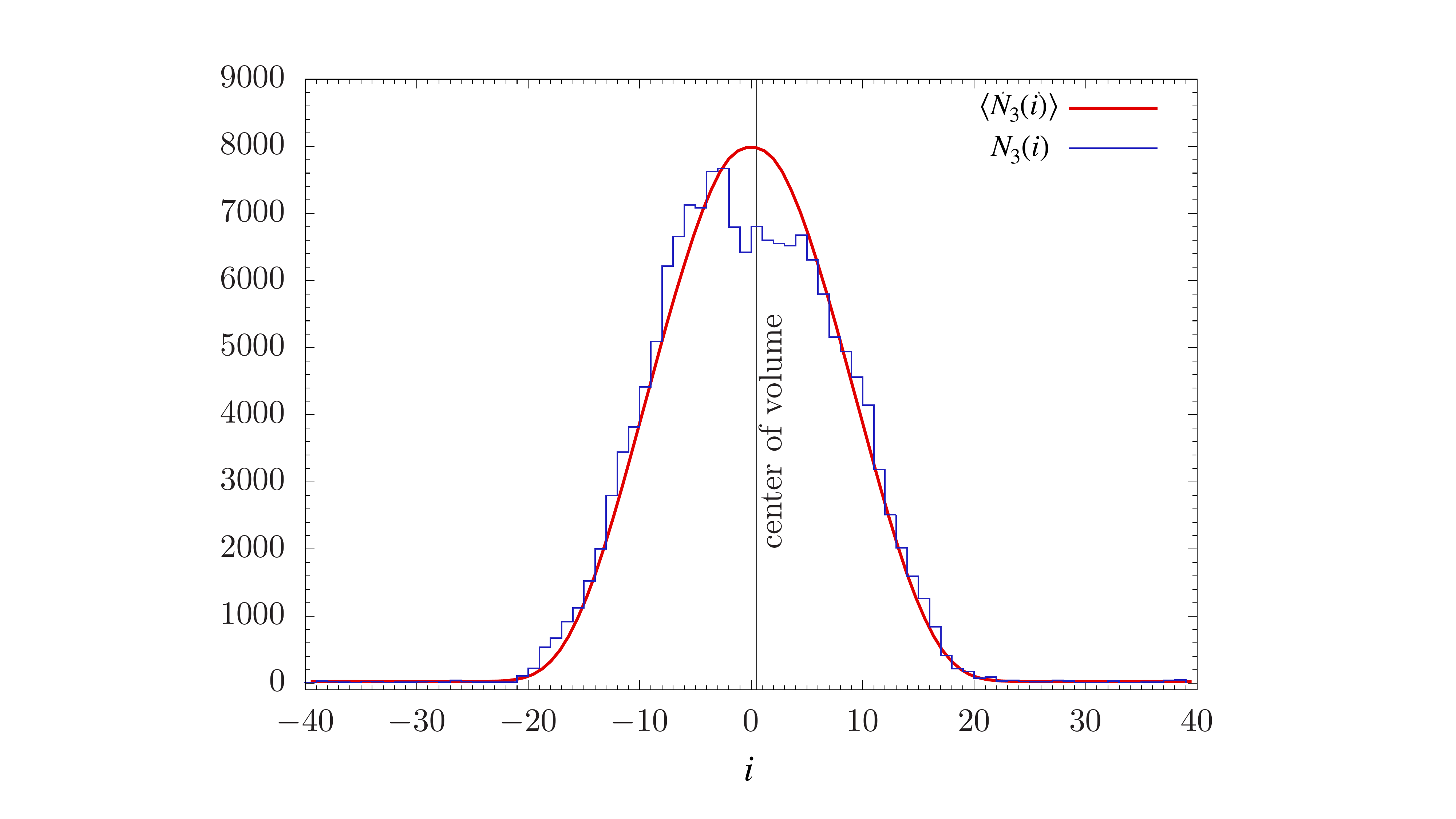}}} 
\hspace{-1.5cm}
{\scalebox{0.55}{\includegraphics{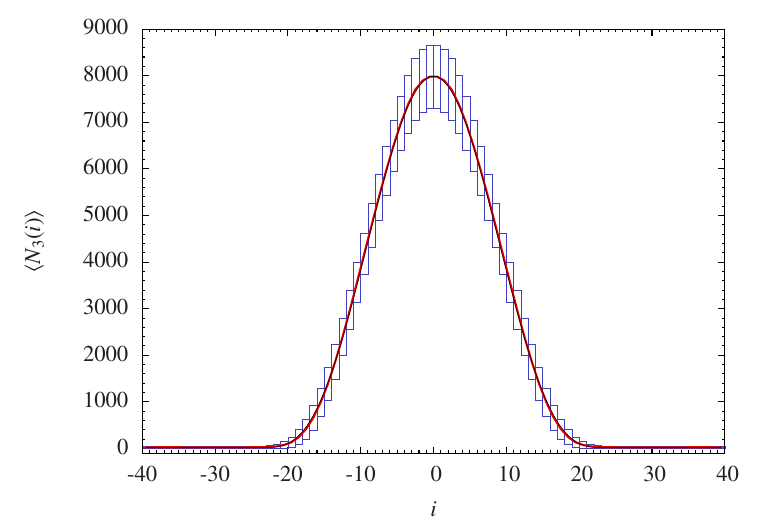}}}}
\vspace{-0.3cm}
\caption{\small Left: single volume distribution $N_3(i)$ for $N_4\! =\! 362.000$ (blue), and averaged volume profile $\langle N_3(i)\rangle$ (red). 
Right: volume profile $\langle N_3(i)\rangle$ (red) and magnitude
$ \sqrt{ \langle \delta N_3(i) \delta N_3(i) \rangle_{N_4}} $ of quantum fluctuations of $N_3(i)$ (blue).}
\vspace{-0.2cm}
\label{fig-am4}
\end{figure}

An example of a typical volume distribution $N_3(i)$, centred around a single peak, is shown in Fig.\ \ref{fig-am4}, left. 
Characteristically, the volume is nonvanishing in a finite time interval, and ``effectively zero'' elsewhere\footnote{i.e.\ has minimal size $N_3(i)\! =\! 5$ compatible with 
a simplicial manifold of topology $S^1\!\times S^3$}.  
In computing the volume profile, the individual distributions have been aligned such that their peak (``centre of volume") always lies at time $i\! =\! 0$, cf.\ Fig.\ \ref{fig-am4}, right.
For sufficiently large $N_4$, the measured volume profile in the region where it is non-minimal follows the functional form 
\begin{equation}
\label{n3av}
\langle N_3(i) \rangle_{N_4} \propto 
N_4 \;  \text{\footnotesize $\frac{1}{ \omega N_4^{1/4}}$} \cos^3\Big(\, \text{\footnotesize $\frac{i}{\omega N_4^{1/4}}$} \Big)
\end{equation}
with great accuracy, where the subscript $N_4$ indicates that the average is computed in the fixed-volume ensemble. 
A closely related observable is the correlator of the fluctuations $\delta N_3(i)\! :=\! N_3(i_1)\! -\! \langle N_3(i_1) \rangle$ of the three-volume,
which is found to behave like 
\begin{equation}
\label{n3n3}
\langle \delta N_3(i_1)\, \delta N_3 (i_2)\rangle_{N_4} 
=  \Gamma \, N_4 \,
{\cal F}\Big(\, \text{\footnotesize $\frac{i_1}{\omega N_4^{1/4}}$ , $\frac{i_2}{\omega N_4^{1/4}}$} \Big), 
\end{equation}
for a universal function $\cal F$, 
where the constants $\omega$ and $\Gamma$ depend on the coupling constants $(k_0,\Delta )$.
 
The volume profile shows a perfect finite-size scaling in the region where the volume is larger than the cutoff size $N_3(i)\! =\! 5$.
In line with relation (\ref{n3av}), its height scales like $N_4^{3/4}$ and its time extension like $N_4^{1/4}$. This constitutes highly nontrivial evidence that the global scaling
behaviour of the dynamically generated quantum universe is that of an extended, four-dimensional spacetime. 
This is demonstrated by introducing the scaling variables 
\begin{equation}
s_i := \text{\footnotesize $\frac{i}{N_4^{1/4}}$}, \quad n_3(s_i) := \text{\footnotesize $\frac{N_3(i)}{N_4^{3/4}}$},
\end{equation}
and comparing the curves for $\langle N_3(i) \rangle_{N_4}$ and 
$ \sqrt{N_4} \,  \langle \delta N_3(s_i)\, \delta N_3(s_j)\rangle_{N_4}$ 
for different four-volumes $N_4$. They collapse to universal curves \cite{Ambjorn2008,review1}
\begin{align}
\langle n_3(s) \rangle & \propto \text{\footnotesize $\frac{3}{4\omega }$} \, \cos^3 \Big(\text{\footnotesize  $\frac{ s}{\omega}$} \Big), \label{curvmatch1}\\
\quad \sqrt{N_4}\, \langle \delta n_3(s_i)\, & \delta n_3(s_j)\rangle \propto  
\Gamma\;
{\cal F} \Big(\text{\footnotesize $\frac{s_i}{\omega},\frac{s_j}{\omega}$} \Big). \label{curvmatch2}
\end{align}
However, these findings allow for a sharper conclusion, beyond
identifying a quantum universe that is macroscopically four-dimensional. Namely, one can
match its overall \textit{shape} -- the volume profile (\ref{n3av}) -- to that of a well-known solution of the classical Einstein equations! 
The observed behaviour of (\ref{n3av}) can be derived from the effective action 
\begin{equation}
\label{seffdisc}
S_{\rm eff}[k_0,\Delta] = \frac{1}{\Gamma}\, \sum_i  \Big(
\text{\footnotesize $\frac{\big(N_3({i+1}) -N_3(i)\big)^2}{ N_3({i} \big)}$ }\!\! + \delta \; N_3^{1/3}(i) \Big)
\end{equation}
for the discrete three-volume, where $\delta$ is a constant depending
on $k_0$ and $\Delta$.
For sufficiently large $N_4$ and using the rescaled volumes $n_3$ and a continuum (Euclidean) proper time $\tau\!\propto\! 1/N_4^{1/4}$,
this action can be rewritten as the inte\-gral expression
\begin{equation}
\label{seffcont}
S_{\rm eff}[k_0,\Delta] = \Big( \text{\footnotesize $\frac{\omega_0}{\omega}$} \Big)^{\! 2} \text{\footnotesize $\frac{\sqrt{N_4}}{\Gamma}$} 
\int d\tau \, \Big(  \text{\footnotesize $\frac{ \dot{n}_3^2(\tau)}{ n_3(\tau)}$} + \delta_0 \, n_3^{1/3}(\tau) \Big), \quad
\int  d\tau \, n_3(\tau) = V_4,
\end{equation}  
where $\delta_0\! =\! 9 (2\pi^2)^{2/3}$ and $\omega_0 = (3/(8\pi^2))^{1/4}$. 
In order to change the constant $\delta$ in the action (\ref{seffdisc}) to the constant $\delta_0$ in (\ref{seffcont}),   
the edge length $a_t$ in the time direction has been rescaled according to
\begin{equation}
\label{eq:arescale}
a_t \!=\! a \,\to\, a_t\! =\! \Big(\frac{\omega_0}{\omega}\Big)^{\! 4/3} a.
\end{equation}
The anisotropic nature of the CDT lattices allows for such a finite relative scaling between the
time- and spacelike edges. 

The action (\ref{seffcont}) is remarkable, since it has the form of a minisuperspace action, written in terms of
the spatial three-volume $n_3(\tau)$ rather than the Friedmann scale factor $a(\tau)\propto n_3(\tau)^{1/3}$. It makes it possible to interpret the measured volume
profile as that of a four-dimensional \textit{Euclidean de Sitter space} -- a round four-sphere -- whose line element is given by
\begin{equation}
\label{lineel}
ds_\mathrm{dS}^2=d\tau^2 + c^2 \cos^2(\tau/c)\, d\Omega^2_{(3)},
\end{equation}
where $d\Omega^2_{(3)}$ denotes the line element of the unit three-sphere, and where
\begin{equation}
\label{n3cont}
n_3(\tau) = 2\pi^2 c^3 \cos^3 (\tau/c)
\end{equation}
is the classical solution obtained by varying the action (\ref{seffcont}), subject to the constraint $\int  d\tau \, n_3(\tau)\! =\! V_4$ on the four-volume. 
The constant $c$ appearing in eqs.\ (\ref{lineel}) and (\ref{n3cont}) is related to the volume of the four-sphere by
$V_4=\tfrac{8}{3} \pi^2 c^4$.  

In addition to matching the (expectation value of the) shape of the quantum universe to that of a classical de Sitter space, also the
behaviour of the fluctuations of the three-volume around this solution, captured by the relations (\ref{n3n3}) and (\ref{curvmatch2}), can be
mapped to a semiclassical analysis, at least for the low-lying part of the fluctuation spectrum \cite{Ambjorn2008}.

One should appreciate the stark difference between the derivation of the de Sitter behaviour of the scale factor
(equivalently, the three-volume) in CDT lattice quantum gravity and in Euclidean quantum cosmology \`a la Hartle-Hawking. 
In the latter one starts from the Einstein-Hilbert action (\ref{eq:contact}), and by fiat imposes spatial homogeneity and isotropy (conditions
meant to hold for cosmology on \textit{large} scales), which reduces all of gravity's $g_{\mu\nu}(x)$ to the scale factor (``conformal mode'') $a(\tau)$. 
One then evades the conformal divergence of the Euclidean 
cosmological path integral, due to the unboundedness of the kinetic part of the minisuperspace action, by adopting a suitable rotation 
of the conformal mode `by hand'. 

By contrast, the lattice formulation does not rely on ad hoc assumptions about global symmetries or on a special treatment of the conformal mode in the Wick rotation.
Instead it provides a bona fide derivation within full-fledged quantum gravity of the minisuperspace result for the global scale factor, in the sense of expectation values, 
provided one identifies
\begin{equation}
\label{eq:const}
\Big( \frac{\omega_0}{\omega} \Big)^{\! 2}
\frac{\sqrt{N_4}}{\Gamma} = \frac{\sqrt{V_4}}{24 \pi G} =
\frac{\sqrt{6}}{24 \Lambda G},
\end{equation}
where $V_4$ is the volume of the four-sphere that solves the Euclidean 
Einstein equations with cosmological constant $\Lambda$.
Note that selecting the three-volume at a given cosmological proper time $\tau$ as a quantum observable is tantamount to ``integrating out'' all other
degrees of freedom in the nonperturbative path integral. Unlike in quantum cosmology, where one removes all local degrees of freedom at the outset, \textit{they are 
still present in the nonperturbative formulation} and are not constrained a priori to be ``close to" a metric space described by the classical line element  
(\ref{lineel}). On the contrary, their collective behaviour gives rise to nonperturbative quantum signatures, like the anomalous spectral dimension discussed below,
and provides natural candidates for Planck-scale inhomogeneities in terms of their local curvature properties, which could play the role of purely gravitational seeds of structure.
This opens the door to investigating quantum-cosmological properties beyond the standard, perturbative treatment
around exactly homogeneous and isotropic universes. 

\addcontentsline{toc}{section}{7\hspace{0.3cm} Infrared and ultraviolet limits of CDT}
\section*{7\hspace{0.2cm} Infrared and ultraviolet limits of CDT}

In a conventional lattice field theory, infrared (IR) and ultraviolet (UV) fixed points of the renormalization group are located 
on critical surfaces in the lattice coupling constant space where the 
correlation lengths of suitable observables are infinite. If the bare lattice coupling 
constants are kept fixed, the renormalized (continuum) coupling constants will 
flow to an IR fixed point when the correlation length goes to infinity. If the 
renormalized coupling constants are kept fixed, the bare lattice coupling constants 
will flow to a UV fixed point. In lattice quantum gravity, the bare couplings are $k_0$ and $\Delta$ 
and one can choose the dimensionless coupling constant $\Lambda G$ as 
a continuum renormalized coupling constant \cite{Ambjorn2024a}. Then the expression on the left in eq.\ (\ref{eq:const}) can be read as 
the lattice expression for $\Lambda G$, expressed in terms of $k_0$, $\Delta$ and 
$N_4$. Because of finite-size scaling one can view $N_4^{1/4}$ as a correlation
length, which implies that the critical surface is given by $N_4 = \infty$.

According to eq.\ (\ref{eq:const}), keeping $k_0$ and $\Delta$ constant and taking $N_4\! \to\! \infty$,
$\Lambda G$ will flow to an IR fixed point. Since
$\Gamma$ and $\omega$ will go to constant values (which depend on $k_0,\Delta$), we have
$\Lambda G\! \to\! 0$ as $N_4 \!\to\! \infty$. Using the continuum renormalization group
one can study the flow of $\Lambda G$, and finds that 0 is an infrared Gaussian fixed point
where $\Lambda \to 0$ and $G \to \ell_p^2$, the squared Planck length.
According to eqs.\ (\ref{eq:arescale}) and (\ref{eq:const}), the lattice representation of this is
\begin{equation}
\label{eq:scalev4}
V_4 \propto \Big(\frac{\omega_0}{\omega}\Big)^{4/3}N_4\, a^4 \propto N_4 \ell_p^4
 \qquad \;\;
a \propto \Big(\frac{\omega_0}{\omega}\Big)^{2/3}
\frac{ \sqrt{G}}{\sqrt{\Gamma}} \propto \ell_p .
\end{equation}
Any point $(k_0,\Delta)$ in the interior of the $C_{\rm dS}$ phase in coupling constant space belongs to the critical 
surface related to the IR Gaussian fixed point of $\Lambda G$. While the 
four-volume $V_4 \propto N_4 a^4$ goes to infinity, the lattice spacing $a$ does not 
scale to zero, but is of the order of the Planck length.

Relation (\ref{eq:scalev4}) allows for an estimate of the size of the universes simulated on the 
computer. For typical values $(k_0,\Delta)$ in phase $C_{\rm dS}$, the diameter
will be around 20 Planck lengths for $N_4 \approx 400.000$. It is surprising that 
global features of such small universes are well described by the effective action (\ref{seffcont}).

To locate a UV lattice fixed point one should follow a path $(k_0(N_4),\Delta(N_4))$
such that $\Lambda G$ on the right in eq.\ (\ref{eq:const}) stays constant as $N_4 \to \infty$,
i.e.\ a path along which
\begin{equation}
\omega^2\big(k_0(N_4),\Delta(N_4)\big)\,  
\Gamma\big(k_0(N_4),\Delta(N_4)\big) \propto \sqrt{N_4}
\quad {\rm for} \quad N_4 \to \infty .
\end{equation}
Numerical evidence suggests that this is only possible if the path leads to 
the $A$-$C_{\rm dS}$ phase transition line from inside the de Sitter phase
$C_{\rm dS}$. Remarkably, one can find paths along which 
$\omega^2 \Gamma$ behaves 
precisely as required when one approaches the $A$-$C_{\rm dS}$ line, namely as
\begin{equation}
\label{eq:crit}
\omega^2 \Gamma \propto N_4^\delta, \qquad \delta = 0.54 \pm 0.04.
\end{equation}
This suggests that the $A$-$C_{\rm dS}$ phase transition line can be 
viewed as a UV critical line and that the CDT lattice implementation of the gravitational path integral 
can be used to define a nontrivial continuum theory of quantum gravity. Eqs.\ (\ref{eq:scalev4}) and (\ref{eq:crit}) show that the cut-off $a\to 0$
when one approaches this critical line. However, it is numerically demanding to further sharpen the 
estimate (\ref{eq:crit}), and the fact that $\omega(k_0,\Delta) \to 0$ at the 
$A$-$C_{\rm dS}$ phase transition line causes an interpretational problem since the time extension of the de Sitter universe, 
$\omega(k_0,\Delta)N_4^{1/4}$, should go to infinity when $(k_0(N_4),\Delta(N_4))$ 
approaches a UV fixed point. On the other hand, 
clarifying these issues is only a question of additional Monte Carlo simulations.

\addcontentsline{toc}{section}{8\hspace{0.3cm} Observables of quantum gravity: quantum curvature}
\section*{8\hspace{0.2cm} Observables of quantum gravity: quantum curvature}

To extract physical results from the nonperturbative path integral one must identify and measure \textit{observables}, namely
operators that depend on the dynamical degrees of freedom and obey a suitable lattice analogue of the diffeo\-mor\-phism-invariance of
gravitational observables in the continuum. The latter is realized by the invariance under any discrete relabelling of the elements of the underlying lattice,
such as its vertices or edges. In pure gravity and without any background geometry or distinguished 
reference frames\footnote{provided by (sufficiently classical) boundaries
or matter distributions}, as is the case here, this implies that observables are necessarily nonlocal, often given by spacetime integrals of local quantities.
Since the lattice simulations explore a near-Planckian regime, with a priori unknown and presumably highly nonclassical properties, it is not
immediately clear which observables can be defined and exist as finite operators. One example of a well-behaved observable is the volume profile discussed earlier. 
Given the extent to which nonperturbative formulations of quantum gravity generally struggle to recover \textit{any} aspect of classical gravity, 
it is truly remarkable that the behaviour of the expectation value of the volume profile observable in CDT has been shown to match that of a
classical de Sitter space. 

The difficulty of constructing meaningful observables is illustrated by the quest for a quantum version of \textit{curvature}, a notion central
to understanding spacetime in classical general relativity. The absence of a smooth background (and accompanying local coordinate systems and
tensor calculus) poses an obstacle to defining curvature modelled on the classical Riemann tensor.
There \textit{is} a natural notion of curvature on finite piecewise flat triangulations based on the concept of deficit angles, familiar from Regge calculus \cite{Regge}, 
but unfortunately it diverges in the continuum limit $N_4\!\rightarrow\! \infty$, without an obvious way of how to renormalize it. 

\begin{figure}[t]
\centerline{\scalebox{0.5}{\rotatebox{0}{\includegraphics{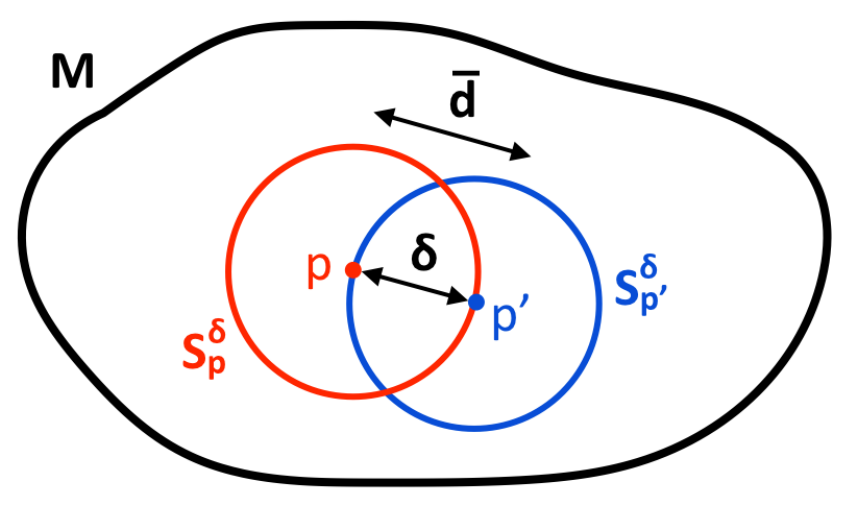}}}}
\caption{Comparing the distance $\bar{d}(S^\delta_p,S^\delta_{p'})$ of two $\delta$-spheres $S_p^\delta$ and
$S^\delta_{p'}$ with the distance $\delta$ of their centres $p$ and $p'$, schematically illustrated for a continuum manifold $M$.}
\label{fig:riccispheresnew}
\end{figure}

Interestingly, a resolution to these issues exists in the form of the so-called \textit{quantum Ricci curvature} \cite{Klitgaard2017}, a notion of curvature applicable to
nonsmooth spaces. It is based on the geometric insight that in the presence of positive Ricci curvature, two nearby geodesic spheres $S_p^\delta$ and
$S_{p'}^\delta$ of equal radius $\delta$ are -- as sets --
nearer to each other than the distance $d(p,p')$ between their respective centres $p$ and $p'$, while they are further apart for negative Ricci curvature. 
In a quantitative implementation of this idea, one sets $d(p,p')\! =\! \delta$ and computes the lattice analogue of the average sphere distance 
\begin{equation}
\bar{d} (S^\delta_p,S^\delta_{p'}):=\frac{1}{vol(S_p^\delta)}\frac{1}{vol(S^\delta_{p'})}\int_{S^\delta_p}d^{3}\! q\,\sqrt{{\det h}}\int_{S^\delta_{p'}} 
d^{3}\! q'\sqrt{\det h'}\ d(q,q')
\label{spheredist}
\end{equation}
in terms of elementary distance and volume measurements on the dynamical triangulations, where $h$ and $h'$ are the induced metrics on the two spheres
(see Fig.\ \ref{fig:riccispheresnew}).
The \textit{quantum Ricci curvature $K_q$ at scale} $\delta$
is then defined in terms of the quotient of distances,
\begin{equation}
\bar{d} (S_p,S_{p'})/\delta =: c_q\, (1-K_q(p,p')),
\label{ricdefine}
\end{equation}
where $c_q$ is a non-universal $\delta$-independent constant and $K_q$ captures the non-constant remainder. 
Due to its quasi-local character, the quantum Ricci curvature must still be integrated over spacetime to yield an observable, or used as a local operator insertion
${\cal O}(x)$ in a diffeomorphism-invariant two-point function of the form
\begin{equation}
G[{\cal O},{\cal O}](r) \! =\! \int_M\!\! d^4x \sqrt{| g(x)|} \int_M\!\! d^4y \sqrt{| g(y)|}\, {\cal O}(x){\cal O}(y)\, \delta (d(x,y)\! -\! r),
\label{2point}
\end{equation} 
suitably normalized. The average quantum Ricci curvature of the dynamically generated quantum geometry in the CDT de Sitter phase as a function of the coarse-graining scale $\delta$
was measured in \cite{Klitgaard2020} and found to be compatible with the behaviour of the same quantity on a classical de Sitter space, adding further evidence
that classical properties can indeed be reproduced by the nonperturbative path integral.

\addcontentsline{toc}{section}{9\hspace{0.3cm} Quantum signature and fractal dimensions}
\section*{9\hspace{0.2cm} Quantum signature and fractal dimensions}

In addition to verifying aspects of a well-defined classical limit in terms of the large-scale behaviour of suitable observables, a key aim
of quantum gravity is to find genuine quantum signatures originating in the Planckian dynamics, beyond perturbative $\hbar$-effects.
A beautiful example of such an observable is the so-called \textit{spectral dimension} $D_S$ of spacetime, which is 
the ``effective" dimension felt by a diffusion process for short diffusion times. It is an
example of a so-called fractal dimension\footnote{Another example is the Hausdorff dimension, which characterizes the volume growth of geodesic balls as a
function of their radius, and can be extracted by setting ${\cal O}(x)\! =\! \mathbf{1}$, the unit operator, in eq.\ (\ref{2point}).}, since it can assume non-integer values on non-classical spaces that are not manifolds but allow for diffusion, such as fractals or graphs.

A well-known early result of CDT lattice quantum gravity is that the spectral dimension does \textit{not} behave classically, but near the Planck scale undergoes 
a continuous ``dynamical dimensional reduction" from the classical, large-scale value of $D_S\! =\! 4$ to a value compatible with 2 \cite{Ambjorn2005}.   
To obtain the (average) spectral dimension, one considers diffusion on individual triangulations $T$ and determines the return
probability $P_T(\sigma)$ of random walkers as a function of the discrete diffusion time $\sigma$. Its eigenvalue in the ensemble for constant volume $N_4$ is given by
\begin{equation}
\langle P_{N_4}(\sigma)\rangle  = \frac{1}{Z_E (k_0,\Delta,N_4)} 
\sum_{T|_{N_4(T) =N_4} }   \text{\footnotesize  $\frac{1}{C_T}$}
\; e^{(k_0+6\Delta)N_0(T)
- \Delta N_4^{(4,1)}(T)}\; P_T (\sigma),
\end{equation}  
where the normalization factor was defined in eq.\ (\ref{eq:partn4}) above. It is expected to have the functional form
\begin{equation}
\langle P_{N_4}(\sigma)\rangle = \sigma^{-D_S/2}\, H \Big( \text{\footnotesize $\frac{\sigma}{N_4^{2/D_S}}$}\Big),
\end{equation}
for some function $H$ satisfying $H(0) > 0$. The spectral dimension $D_S(\sigma)$ is extracted from the leading power-law scaling by defining
\begin{equation}
\label{eq:specdim}
D_S(\sigma) := -2\, \frac{d \ln \langle P_{N_4} (\sigma) \rangle}{d\sigma}
\end{equation}  
for sufficiently small $\sigma$. Its expected behaviour, observed in similar lower-dimen\-sio\-nal and/or purely Euclidean ensembles, is  
an approximately constant behaviour of $D_S(\sigma)$ for $\sigma\! <\! N_4^{2/D_S}$, which then justifies calling (\ref{eq:specdim}) \textit{the} 
spectral dimension of the system. 

\begin{figure}[t]
\vspace{-2cm}
\centerline{\scalebox{0.7}{\rotatebox{0}{\includegraphics{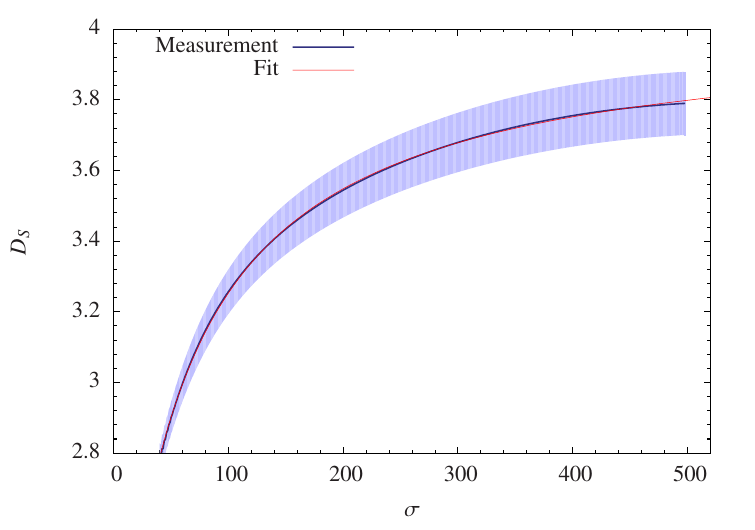}}}}
\vspace{-0.3cm}
\caption[phased]{{\small Spectral dimension $D_S$ as a function of the diffusion time $\sigma$, measured for spacetime volume
$N_4\! =\! 181.000$. The averaged measurements lie along the central curve, together with a best fit
$D_S(\sigma) \! = \! 4.02 -  119/(54+\sigma)$. The two outer curves represent error bars. 
}}
\label{d4s2.2b4}
\end{figure}

However, measuring the return probability $\langle P_{N_4}(\sigma)\rangle$ in CDT lattice quantum gravity
led to the surprising finding that there is no
plateau of constant $D_S(\sigma)$, but instead a \textit{scale-dependent} behaviour not observed before \cite{Ambjorn2005}. This is illustrated
by Fig.\ \ref{d4s2.2b4}, which shows a clear $\sigma$-dependence for small $\sigma$, where $D_S(\sigma)$ increases gradually from around 
2 (more precisely, $D_S(\sigma\!\rightarrow\! 0)\! =\! 1.80\pm 0.25$) to a value compatible with 4 for asymptotically large $\sigma$. 

There are several reasons why this result is very noteworthy. Firstly, the spectral dimension of quantum spacetime at the
Planck scale is a \textit{truly nonperturbative observable}, which characterizes Planckian physics in an invariant way. 
Secondly, it is simply a dimensionless number that should be computable in other formulations of quantum gravity too.
In a field which suffers from a shortage of computable observables, this is very important, because it can be used as a benchmark to compare
different approaches.  

After the discovery of the anomalous behaviour of the spectral dimension in CDT, corroborating evidence for $D_S\! =\! 2$ has
been found in several other approaches, and it has been conjectured to be a universal property of quantum gravity \cite{Carlip2017}.
This formula can therefore be seen as playing a similar role -- albeit in a fully nonperturbative context -- as the Bekenstein-Hawking formula $S=A/4$ 
for the entropy of a black hole as a function of its area, whose origin is semiclassical. Like for the black-hole entropy, the phenomenological consequences of the dimensional
reduction at the Planck scale are currently not known, but uncovering this quantum signature clearly represents progress in the right direction.

\addcontentsline{toc}{section}{10\hspace{0.1cm} State of the art and future challenges}
\section*{10\hspace{0.2cm} State of the art and future challenges}

After decades of research into lattice quantum gravity, there is now an implementation based on causal dynamical triangulations that successfully 
addresses three major technical challenges \cite{review1,review2,Ambjorn2024,Loll2025}:
(i) how to incorporate the dynamical nature of spacetime -- by working with dynamical rather than fixed lattices, (ii) how to incorporate the
Lorentzian character of spacetime while remaining amenable to Monte Carlo simulations -- by working with a set of Lorentzian lattices that allow for a
well-defined Wick rotation, and (iii) how to account for the physical degrees of freedom of gravity without any gauge redundancies -- by working with identical
building blocks and eliminating the ensuing relabelling symmetry.  

The CDT lattice formulation comes with a fully operational computational framework, using well-tested, state-of-the-art computer codes, which are being optimized and 
adapted to new observables on a continuous basis. The Markov chain Monte Carlo (MCMC) simulations typically run with configuration sizes $N_4$ of 
between several hundred thousand and a million building blocks, and are subject to the usual limitations of lattice field theory regarding computing power, lattice size, 
discretization artefacts, finite-size effects and numerical errors. These simulations provide a unique, nonperturbative window on quantum spacetimes of
linear size of the order of 12-20 Planck lengths. To determine their physical properties one measures observables, 
which amounts to conducting numerical MCMC ``experiments" on the regularized path integral ensemble,
and then extrapolates their continuum behaviour through finite-size scaling \cite{review1}. 

A central goal of quantum gravity is to predict new physical phenomena, which cannot be explained by classical gravity and can be verified through observation
or experiment. While quantum effects predicted by perturbative formulations tend to be extremely small by construction and well beyond any range of detection, 
nonperturbative mechanisms provide a potential way out. The fact that CDT lattice gravity has (already) demonstrated the dynamical emergence of a quantum
spacetime with de Sitter-like properties is a promising starting point for connecting its Planckian dynamics to early-universe cosmology, which 
\textit{postulates} that spacetime resembles a homogeneous and isotropic classical de Sitter space.   

Research is under way to establish whether the local properties of CDT's quantum de Sitter space 
are compatible with standard cosmological assumptions or lead to alternative predictions, possibly with phenomenological consequences down the line.
This is a challenging task, both conceptually and computationally, because the nature of the highly quantum-fluctuating Planckian regime is very different from the 
habitat of cosmology, 
that of quantum fields on a fixed, curved background. Key to making this connection is the identification of suitable nonperturbative observables, which 
can be measured reliably with the methods described here and related to typical cosmological observables. Promising candidates, which have already been
tested in lower-dimensional toy models, are diffeomorphism-invariant homogeneity measures and curvature two-point functions, constructed 
along the lines of eq.\ (\ref{2point}). Investigations also include the study of matter, which can be coupled in a straightforward way, and its interaction with quantum geometry. 

Much remains to be understood about the fascinating world of the Planck scale and how it may provide dynamical mechanisms and quantitative explanations 
for the emergence of spacetime and structure in the universe.

\end{document}